\documentclass[aip,
amsmath,amssymb,
preprint,
]{revtex4-2}

\usepackage{graphicx}
\usepackage{dcolumn}
\usepackage{bm}

\usepackage[utf8]{inputenc}
\usepackage[T1]{fontenc}
\usepackage{mathptmx}
\usepackage{etoolbox}

\usepackage{amsmath,gensymb,textcomp,bm,dcolumn,eurosym,array,tabu,multirow,nicefrac,color,graphicx,upgreek}
\newcommand{\bra}[1]{\langle #1 \rvert}
\newcommand{\ket}[1]{\lvert #1 \rangle}

\DeclareMathOperator{\sinc}{sinc}

\newcommand{\pare}[1]{\left( #1 \right)}

\newcommand{\abs}[1]{\left\vert #1 \right\vert}
\newcommand{\cor}[1]{\left[ #1 \right]}

\makeatletter
\def\@email#1#2{%
 \endgroup
 \patchcmd{\titleblock@produce}
  {\frontmatter@RRAPformat}
  {\frontmatter@RRAPformat{\produce@RRAP{*#1\href{mailto:#2}{#2}}}\frontmatter@RRAPformat}
  {}{}
}%
\makeatother
\begin{document}

\preprint{AIP/123-QED}

\title[Sample Classification using Machine Learning-Assisted Entangled Two-Photon Absorption]{Sample Classification using Machine Learning-Assisted Entangled\\ Two-Photon Absorption}
\author{Áulide Martínez-Tapia}
    \email{aulide.martinez@correo.nucleares.unam.mx}
\author{Roberto de J. León-Montiel}%
 \email{roberto.leon@nucleares.unam.mx}
\affiliation{Instituto de Ciencias Nucleares, Universidad Nacional Aut\'onoma de M\'exico, Apartado Postal 70-543, 04510 CDMX, M\'exico
}%

\date{\today}

\begin{abstract}
Entangled two-photon absorption (eTPA) has been recognized as a potentially powerful tool for the implementation of ultra-sensitive spectroscopy. Unfortunately, there exists a general agreement in the quantum optics community that experimental eTPA signals, particularly those obtained from molecular solutions, are extremely weak. Consequently, obtaining spectroscopic information about an arbitrary sample via conventional methods rapidly becomes an unrealistic endeavor. To address this problem, we introduce an experimental scheme that reduces the amount of data needed to identify and classify unknown samples via their electronic structure. Our proposed method makes use of machine learning (ML) to extract information about the number of intermediate levels that participate in the two-photon excitation of the absorbing medium. This is achieved by training artificial neural networks (ANNs) with various eTPA signals where the delay between the absorbed photons is externally controlled. Inspired by multiple experimental studies of eTPA, we consider model systems comprising one to four intermediate levels, whose energies are randomly chosen from four different intermediate-level band gaps, namely: $\Delta\lambda = 10$, 20, $30$, and $40$ nm. Within these band gaps, and with the goal of testing the efficiency of our artificial intelligence algorithms, we make use of three different wavelength spacing $1$, $0.5$ and $0.1$ nm. We find that for a proper entanglement time between the absorbed photons, classification average efficiencies exceed 99$\%$ for all configurations. Our results demonstrate the potential of artificial neural networks for facilitating the experimental implementation of eTPA spectroscopy.

\end{abstract}

\maketitle

\section{\label{sec:level1}INTRODUCTION}

Entangled two-photon absorption spectroscopy belongs to a group of quantum-enhanced techniques that hold promise for extracting unique information about the energy dynamics and electronic structure of complex molecular systems. \cite{schlawin_book} Notably, during the past three decades, time and frequency correlations of entangled photon pairs have provided us with the ability to observe non-trivial quantum effects, such as the linear dependence of the two-photon absorption rate as a function of the photon flux that illuminates a sample.\cite{Javanainen-1990, dayan2005, Lee-2006} They have also been fundamental in the theoretical prediction of fascinating quantum-correlation-enabled phenomena, including the process of entanglement-induced two-photon transparency, \cite{fei1997,guzman2010} the excitation of usually forbidden atomic transitions, \cite{ashok2004} the manipulation of quantum pathways of matter, \cite{roslyak2009-1, raymer2013, schlawin2016, schlawin2013, schlawin2017} and the control of molecular processes. \cite{shapiro2011, shapiro_book}

Time-frequency entangled photons exhibit temporal and spectral characteristics that allow for measurements with high spectral and temporal resolution, something that it is not achievable with conventional methods. \cite{Zhang2022, Zhang:24} More specifically, eTPA spectroscopy allows for probing intermediate single-photon excitation states with broadband photons, while maintaining a high frequency resolution within the two-photon excitation manifold. \cite{schlawin2013} It is interesting to note that this unique feature of time-frequency correlated photons appears, particularly, when these are generated by means of continuous-wave pumping; \cite{Schlawin-2018} thus showing that one can indeed replicate pulsed laser experiments using continuous-wave-pumped entangled-photon sources. \cite{Harper2023} The combination of these properties, together with the possibility of exciting multiphoton processes at low intensities, \cite{Dorfman(2016)} has made entangled photon pairs especially attractive for their use in quantum-enhanced molecular spectroscopy. \cite{saleh, kojima2004, nphoton, roberto_spectral_shape, oka2010, villabona_calderon_2017, Varnavski2017, oka2018-1, oka2018-2, svozilik2018-1, svozilik2018-2, burdick2018, RobertoTemperatureControlled, Mukamel2020roadmap, Mertenskotter:21, villabona2020, parzuchowski2021, tabakaev2021, landes2021, samuel2022, tabakaev2022, Áulide2023, FSchlawin2024} However, there is general agreement in the eTPA community that experimental two-photon absorption signals, particularly those obtained from molecular solution samples, are extremely weak. \cite{raymer2021entangled, landes2021quantifying, raymer2021, mikhaylov2022, cushing2022, Landes2024arxiv} An immediate consequence of this result is that obtaining spectroscopic information about an arbitrary sample using eTPA together with conventional data-handling methods rapidly becomes an extremely complicated, if not unrealistic, task.

In this work, we introduce an experimental scheme designed to address the primary challenge of experimental eTPA, namely the extremely weak signals that can be retrieved from molecular samples. Our proposed method leverages a machine learning (ML) protocol to extract information about the electronic structure of an absorbing sample, specifically the number of intermediate levels that participate in its two-photon excitation. This is managed by training artificial neural networks (ANNs) with multiple model-based eTPA signals recorded only as a function of an external delay between the two correlated photons that are absorbed by the sample. We make use of experimentally attainable external delay values, as well as two-photon bandwidths, \cite{samuel2022} to demonstrate the possibility of a rapid classification of (arbitrary) samples via eTPA assisted by artificial intelligence.

\begin{figure*}
\centering
\includegraphics[width = 16cm]{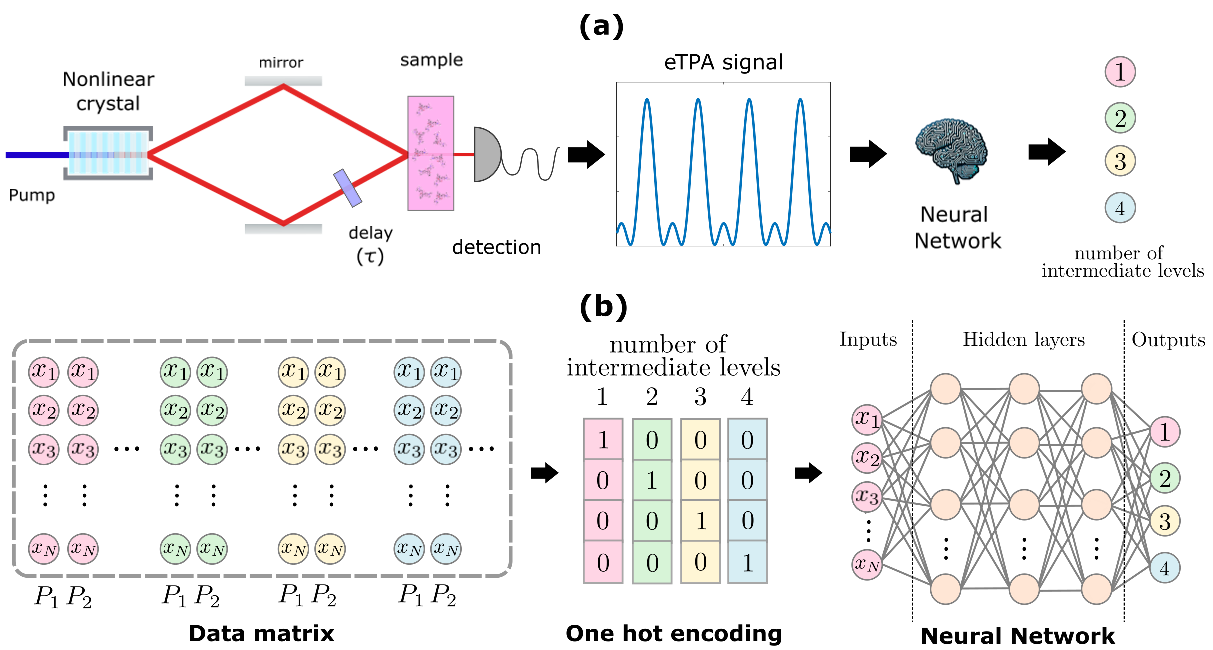}
\caption{Sample classification using artificial-intelligence-assisted eTPA. (a) Our proposed technique makes use of a typical eTPA transmission experiment, \cite{villabona_calderon_2017,samuel2022,cushing2022} which comprises entangled photons pairs produced by spontaneous parametric down-conversion (SPDC). The photons interact with an arbitrary sample that experience a two-photon absorption process. The eTPA signal is recorded as a function of an external delay between the twin photons, which is fed to a model-based trained artificial neural network that returns the number on intermediate levels present in the two-photon absorption process. (b) Artificial intelligence protocol. First we create the data set with the theoretical model of Eq. (\ref{P(tau)}) for four hypothetical systems with a maximum of four intermediate levels. We then use one hot encoding to label the data-set and create a supervised neural network. Finally the neural network classifies the signals according to the number of intermediate levels that participate in the two photon absorption process.}
\label{fig:Fig1}
\end{figure*}

The proposed technique is sketched in Fig. \ref{fig:Fig1}(a). We consider a standard entangled two-photon source consisting of a nonlinear crystal pumped by a continuous-wave laser source. The photon pairs---generated by means of spontaneous parametric down conversion (SPDC)---interact with an arbitrary sample in which two-photon absorption takes place. In order to reveal the information about the intermediate levels that contribute to the two-photon excitation of the sample, a time delay ($\tau$) between frequency-correlated photons is introduced. The eTPA signal is then recorded as a function of this temporal delay. As a proof of principle, we consider four different samples, identified by the number of intermediate states that participate in the two-photon excitation process. Our artificial intelligence protocol [depicted in Fig. \ref{fig:Fig1}(b)] is trained with multiple eTPA signals computed from a general theoretical model of eTPA (see below). We created a database for four hypothetical molecular systems. Then, we use one hot encoding to label the data set and design the supervised neural network shown on the right-hand side of Fig. \ref{fig:Fig1}(b). It is worth pointing out that, as we will show next, our machine learning-based method is capable of dramatically reducing the amount of data that is needed to identify and classify unknown samples via their electronic structure. This shows the potential of the proposed technique for helping current efforts that aim to experimentally demonstrate true eTPA-based imaging and spectroscopy. 

\section{Theoretical Model}
Our proposed method considers a typical eTPA transmission-based experiment, \cite{samuel2022} where pairs of frequency-correlated photons interact with an arbitrary sample after an external delay between them is introduced [see Fig. \ref{fig:Fig1}(a)]. The flux of photon pairs leaving the sample is expected to change, with respect to the incident flux, as a result of the eTPA process. Previous authors have demonstrated that changes in the outgoing flux strongly depend on the delay, $\tau$, between photons. More importantly, the signal has been shown to carry spectroscopic information that can be retrieved by means of a Fourier transform. \cite{Saleh-1998,RobertoTemperatureControlled} It is worth remarking that retrieval of the desired information, namely the electronic structure, can only be obtained by adding another degree of freedom, namely changes in the eTPA signal as a function of the correlation (entanglement) time between photons, \cite{Saleh-1998} the temperature of the crystal that generates them, \cite{RobertoTemperatureControlled} or the pump frequency. \cite{Mertenskotter:21} Evidently, this suggests that eTPA spectroscopy might require large measurement sessions. We remark that there exists another method for extracting the eTPA, which ideally has a zero-background signal. This technique is based on the detection of the fluorescence produced by the absorption of the correlated photon pairs, see References \cite{Varnavski2017, villabona_calderon_2017, Landes2024arxiv} for details.

To model the quantum light-matter interaction, we assume that a two-photon field, $\ket{\psi}$, excites the absorbing medium from an initial state $\ket{g}$ (with energy $\epsilon_{g}$) to a doubly-excited final state $\ket{f}$ (with energy $\epsilon_{f}$), via non-resonant intermediate states $\ket{j}$ (with energy $\epsilon_{j}$). The transitions are dipole-mediated and thus the interaction Hamiltonian is taken to be
$\hat{H}(t)=\hat{d}(t)\hat{E}^{(+)}(t)+ \text{H.c.}$ (H.c. stands for the Hermitian conjugate), where $\hat{d}(t)$ is the dipole-moment operator, and $\hat{E}^{(+)}(t)$ is the positive-frequency part of the electric field operator. The field operator is written as $\hat{E}^{(+)}(t)=\hat{E}^{(+)}_{s}(t)+\hat{E}^{(+)}_{i}(t)$, with $\hat{E}^{(+)}_{s,i}(t)$ denoting the signal ($s$) and idler ($i$) fields:

\begin{equation}
    \hat{E}^{(+)}_{s,i}(t) = \int\,d\omega_{s,i}\sqrt{\frac{\hslash\omega_{s,i}}{4 \pi \epsilon_0 c A}}\hat{a}(\omega_{s,i})e^{-i\omega_{s,i}}\label{E+s,i},
\end{equation}

\noindent
with $c$ denoting the speed of light in vacuum, $\epsilon_0$ the vacuum permittivity, $A$ the effective area where the interaction occurs, and $\hat{a}(\omega_{s,i})$ the annihilation operator for the photonic mode of the signal and idler photons, respectively. For simplicity, we do not write the spatial shape and polarization of each mode.
Then, using second-order time-dependent perturbation theory, we write the probability of two-photon excitation of the medium as \cite{nphoton, roberto_spectral_shape, RobertoTemperatureControlled}

\begin{equation}
    \label{TPAProb}
    P_{g\rightarrow f}=\abs{\frac{1}{\hslash^2}\int_{-\infty}^{\infty} \,dt_2 \int_{-\infty}^{t_2} \,dt_1 M_{\hat{d}}(t_1,t_2)M_{\hat{E}}(t_1,t_2)}^2,
\end{equation}

\noindent
where $M_{\hat{d}}$ and $M_{\hat{E}}$ are terms describing the behavior of the absorbing medium and the two-photon field, respectively. These are given by

\begin{equation}
    M_{\hat{d}}(t_1,t_2) = \sum_{j=1} D^{(j)} e^{-i(\epsilon_j-\epsilon_f) t_2}  e^{-i(\epsilon_g-\epsilon_j) t_1} ,\label{Md}
\end{equation}

\begin{equation}
M_{\hat{E}}(t_1,t_2) =\bra{\text{vac}}\hat{E}_2^{(+)}(t_2)\hat{E}_1^{(+)}(t_1)\ket{\psi}
+ \bra{\text{vac}}\hat{E}_1^{(+)}(t_2)\hat{E}_2^{(+)}(t_1)\ket{\psi}.\label{ME}
\end{equation}

\noindent
The term describing the medium's structure, Eq. (\ref{Md}), depends on the transition matrix elements of the dipole-moment operator, $D^{(j)}=\bra{f}\hat{d}\ket{j}\bra{j}\hat{d}\ket{g}$, and shows that the excitation of the medium occurs through the intermediate states $\ket{j}$. For the sake of simplicity, we have assumed that the lifetime of the final state is much larger than the lifetime of the intermediate states. On the other hand, the term describing the two-photon field, Eq. (\ref{ME}), tells us that only one photon of each field contributes to the two-photon excitation. Note that the two terms in Eq. (\ref{ME}) describe the two possible interaction pathways through which the sample is excited. 
To complete the eTPA model, we define the initial continuous-wave-pumped, two-photon state as \cite{roberto_spectral_shape,RobertoTemperatureControlled}

\begin{equation}
        \ket{\psi} = \left( \frac{T_e}{\sqrt{\pi}} \right)^{1/2} \int_{-\infty}^{\infty}\int_{-\infty}^{\infty} \,d\omega_s\,d\omega_i\delta(\omega_p-\omega_s-\omega_i)\sinc\cor{T_e \pare{\omega_{i}-\omega_{s}}}e^{-i\omega_i\tau}\hat{a}^{\dagger}_s(\omega_s)\hat{a}^{\dagger}_i(\omega_i)\ket{0}, \label{JSI}
\end{equation}
where $\omega_j$ ($j = p, s, i$) stands for the frequencies of the pump, signal and idler fields, respectively. $T_e = (N_{s} - N_{i})L/4$ is the correlation or entanglement time, with $L$ denoting the length of the nonlinear crystal that generates the photon pairs, and $N_{s,i}$ the inverse group velocities of the signal and idler modes, respectively. Note that Eq. (\ref{JSI}) includes a phase function, which represents the external delay, $\tau$, introduced between the signal and idler photons. 

By substituting Eqs. (\ref{Md})-(\ref{JSI}) in Eq. (\ref{TPAProb}), it is straightforward to find that the eTPA probability can be written as
\begin{widetext}
\begin{equation}
    P_{g\rightarrow f}(\tau) = \frac{\abs{\delta\left(\frac{\Delta_+}{2\pi}\right)}^2}{4\pi\hslash^2\epsilon_0^2 c^2 A^2} \frac{\omega_{i}^{0}\omega_{s}^{0}}{T_e} \abs{ \sum_{j=1} D^{(j)} \Biggl\{ \frac{1-e^{-i\left[ \epsilon_j -\omega_{i}^{0} \right](2T_e-\tau)}}{\epsilon_j - \omega_{i}^{0}} + \frac{1-e^{-i\left[ \epsilon_j -\omega_{s}^{0} \right](2T_e+\tau)}}{\epsilon_j - \omega_{s}^{0}} \Biggr\} }^2,\label{P(tau)} 
\end{equation}
\end{widetext}
where $\Delta_{+}=\pare{\omega_{p}-\epsilon_{f}}/2$, and $\omega_{s,i}^{0}$ correspond to the central frequencies of the signal and idler photon wavepackets, respectively. For the sake of simplicity, we have displaced the energy of the ground state to zero, i.e., $\epsilon_{g}=0$. Moreover, we have assumed the condition $\omega_{s}^{0}+\omega_{i}^{0}=\epsilon_{f}$, which guarantees that the two-photon field is completely resonant with the $\ket{g}\rightarrow\ket{f}$ transition of the hypothetical molecular sample.

\begin{figure}
    \centering
    \includegraphics[width = 14cm]{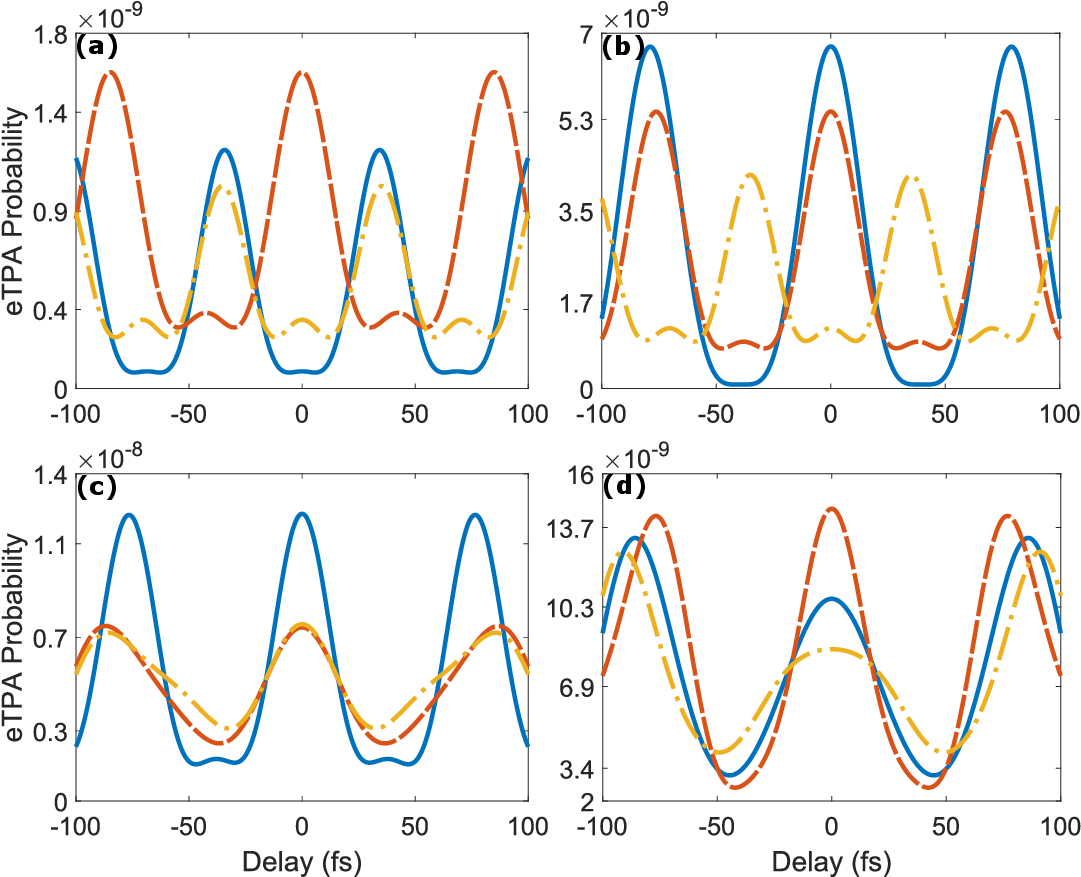}
    \caption{Examples of eTPA signals obtained via the model in Eq. (\ref{P(tau)}), considering: (a)one, (b)two, (c)three, and (d)four intermediate levels. Individual panels show three different signals that are obtained when randomly selecting the energy values for each level configuration. Note the strong similarity between signals comprising different number of intermediate states. The evaluation of Eq. (\ref{P(tau)}) is performed assuming degenerate photon pairs, where $\omega_{s}^{0}=\omega_{i}^{0}=\omega_{0}=2\pi c/(810\;\text{nm})$. The interaction area is set to $A=10\;\mu\text{m}^{2}$, and the entanglement time is taken to be $T_{e}=63\;\text{fs}$.
    }
    \label{fig:Fig2}
\end{figure}

Equation (\ref{P(tau)}) allows us to model multiple eTPA signals (some examples are shown in Fig. \ref{fig:Fig2}), which are later used to train our ML pattern recognition protocol. Experimentally, the eTPA signals may be obtained by changing the external delay and recording the rate of coincidence in transmission, i.e., how many correlated photon pairs are lost during the eTPA process. The plots shown in Fig. \ref{fig:Fig2} correspond to model systems with (a) one, (b) two, (c) three, and (d) four arbitrarily chosen intermediate levels. For simplicity, we have assumed that photon pairs are degenerate, i.e., $\omega_{s}^{0}=\omega_{i}^{0}=\omega_{0}$, and that the intermediate levels are near-resonant to their central frequencies, $\omega_{0}=2\pi c/(810\;\text{nm})$, which is the condition where efficient eTPA can be expected. \cite{upton_2013,burdick2018} The evaluation of Eq. (\ref{P(tau)}) is performed assuming an interaction area of $A=10\;\mu\text{m}^{2}$, and an entanglement time of $T_{e}=63\;\text{fs}$. Note that this is the latest (and shortest) entanglement time used in experimental eTPA investigations. \cite{Varnavski2017} In all configurations, the eTPA shows a non-trivial behavior as a function of the delay, which makes it clear that there is not a straightforward way to differentiate between them by just observing the signal. In a realistic scenario, if we were to analyze multiple signals, without a priori knowledge of the samples, the task of sample classification becomes an extremely complicated, if not impossible, endeavor. This motivates the use of a machine learning classification protocol.

\section{Machine-learning-assisted eTPA Spectroscopy}

The goal of our artificial intelligence-assisted technique is to identify the number of intermediate levels present in an eTPA signal. As a proof of principle of the method, we make use of computationally generated eTPA signals via Eq. (\ref{P(tau)}). Inspired by previous experimental work, we consider four hypothetical molecular species, each with different number of intermediate levels. The delay-dependent eTPA signals are then labeled according to the number of intermediate levels that participate in the two-photon transition. Once ready, the data-set is injected into a pattern-recognition neural network for training, validation, and testing.

As a mean on building up the data-set, we computed five hundred different eTPA signals for each of the four classes. This creates a data-set of two thousand eTPA signal samples. It should be noted that we considered the first class as having one intermediate level, the second one having two intermediate levels, and so on, so the fourth class has four intermediate levels. To show the effectiveness of our method, we start with the case where intermediate-state levels (in terms of wavelength) are randomly chosen from a set of values $\lambda = (835, 845)$ nm, which corresponds to a band of intermediate states of $\Delta\lambda = 10$ nm centered at $840$ nm. The levels are uniformly distributed for three different cases, where the minimum distance between levels is $1$ nm, $0.5$ nm, and $0.1$ nm, respectively. The latter case is expected to be the most complex, as it includes many more combinations than the first two. In fact, this case is considered to emulate the most common situations in which we have no information \textit{a priori} about the intermediate levels of the samples. We then use these signals to create the input matrices. Each column of the input will have five hundred elements. Note that the number of divisions corresponds to the sampling of the eTPA signal in a time interval of $\tau = (-100, 100)$ fs. It is worth pointing out that the time divisions are well within what can be experimentally managed. Specifically, the 500-element division requires a time-step of 0.4 ps.

For the output matrix, we use the \textit{one hot encoding} codification method, which consists of creating column vectors for each unique value that exists in the categorical variable that we are coding [see Fig. 1 (b)]. We then mark the column corresponding to the value present in each record with a 1, leaving the other columns with a value of 0. In our case, for the variable \textit{number of intermediate levels of the system}, one hot encoding creates a column vector with 4 entries, one for each of the four different samples. This way each eTPA signal is assigned to the column corresponding to the number of intermediate levels it has, the first column for one intermediate state, the second column for two intermediate states, and so on. By using this method, each record is represented by a binary vector that indicates the presence or absence of each categorical value. This codification method avoids the possibility that the algorithm may misinterpret the numerical values assigned by the label encoding. 

\subsection{Results}\label{subsec1}

In order to identify the number of intermediate states that participate in the two-photon excitation of arbitrary samples, our protocol takes advantage of the self-learning properties of artificial neurons. The architecture of our neural networks comprises sigmoid neurons for the hidden layers, whereas softmax neurons are used for the output layer. The training is performed using the scaled conjugate gradient back-propagation algorithm, \cite{MOLLER1993525} which optimizes the performance in a direction that minimizes the cross-entropy. The cross-entropy is used as the loss function due to its proven effectiveness in classification tasks.\cite{Shore1981IEEE} The data-set was divided into training with $70\%$ of the samples, validation with $15\%$, and test with the remaining $15\%$. The training set is used for the network to \emph{learn} the different shapes of eTPA signals. The training continues as long as the network shows an improvement in the validation set. Finally, the test set provides a completely independent measure of the network's accuracy, as the data contained in that set is never seen by the network during the training process. 

Table \ref{tab:table1} shows the results of the best ANNs used for an intermediate-level band of $\Delta\lambda=10$ nm [in a set of possible wavelengths of $(835, 845)$] for an entanglement time of $T_e= 63$ fs. We found the mean classification efficiencies, $E_m$, with their corresponding standard deviations, $\sigma$, to be $99.82\pm0.61\%$, $99.88\pm0.48\%$, and $99.71\pm1.18\%$ for the three different steps 1, 0.5 and 0.1 nm, respectively. Note that in order to find the mean efficiency, we run the classification algorithm one hundred times, considering different initial random weights and bias each time. We finally average over the efficiency of each NN (which contains a single hidden layer with five sigmoid neurons). The efficiency corresponds to the success rate of correctly classifying the eTPA signal as produced by a sample with one, two, three or four intermediate states.

\begin{table*}[t!]
    \caption{\label{tab:table1} Summary of the overall performance of our neural networks. The first and second column show the four different intermediate-level band sizes, namely $(\Delta\lambda = 10,\;20,\;30,\;40)$ nm. The third column shows the intermediate-level spacing within the bands. The fourth and fifth columns contain the mean ANN's classification efficiency, and its standard deviation, for an entanglement time of $T_e= 63$ fs. The last two columns corresponds to the mean ANN's classification efficiency, and its standard deviation, assuming an entanglement time of $T_e= 7.16$ fs. We tested our algorithms using this entanglmente time, because it has been shown that one can experimentally achieve such ultrabroadband photon pairs in the context of continous-wave pumped SPDC sources. \cite{Nasr2008} Note that the mean classification efficiency of our neural networks is obtained by averaging the efficiency of one hundred different ANNs.}
    \begin{ruledtabular}
        \begin{tabular}{c c c c c c c}
            \multicolumn{3}{c}{\textbf{Intermediate level bands}}        &\multicolumn{2}{c}{ $\bm{T_e= 63}$ \textbf{fs}} & \multicolumn{2}{c}{ $\bm{T_e= 7.16}$ \textbf{fs}}\\
            range     (nm)       & $\Delta\lambda_{\textit{int}}$ (nm)  & step (nm)    & $E_{m} (\%)$ &  $\sigma (\%)$                & $E_{m} (\%)$  &  $\sigma (\%)$\\
            
            \hline
            \multirow{3}{*}{$(835,845)$} &                    &  1       &  99.82       &  0.61   & 99.64  & 2.77 \\
                                         &     10           &  0.5     &  99.88       &  0.48   & 99.53  & 2.88 \\
                                         &                    &  0.1     &  99.71       &  1.18   & 99.48  & 3.16 \\
            \multirow{3}{*}{$(830,850)$} &                    &  1       &  99.52       &  1.02   & 98.70  & 4.92 \\
                                         &     20           &  0.5     &  99.45       &  2.11   & 99.62  & 0.83 \\
                                         &                    &  0.1     &  99.37       &  1.50   & 98.99  & 4.12 \\
            \multirow{3}{*}{$(825,855)$} &                    &  1       &  84.74       &  5.47   & 99.22  & 2.32 \\
                                         &    30            &  0.5     &  84.10       &  6.46   & 99.37  & 1.10 \\
                                         &                    &  0.1     &  85.06       &  6.02   & 99.18  & 3.82 \\
            \multirow{3}{*}{$(820,860)$} &                    &  1       &  66.66       &  5.06   & 99.26  & 3.09 \\
                                         &    40            &  0.5     &  68.36       &  4.51   & 99.59  & 2.06 \\
                                         &                    &  0.1     &  68.51       &  3.94   & 99.35  & 3.61 \\
        \end{tabular}
    \end{ruledtabular}
\end{table*}

To further explore the performance of our method in more complex scenarios, we applied the protocol for three more cases which have broader intermediate-level bands, namely $ (830, 850)$ nm, $ (825, 855)$ nm, and $ (820, 860)$ nm. These correspond to bandwidths of $\Delta\lambda = 20$, $30$, and $40$ nm, respectively. As described above, we consider three scenarios where the accesible levels are sampled using a resolution of 1, 0.5 and 0.1 nm. The mean efficiencies, and corresponding standard deviations, are also reported in Table \ref{tab:table1}. Note that the best average efficiencies for the maximum sampling resolution (0.1 nm) are $99.37\pm1.50\%$, $85.06\pm6.02\%$, and $68.51\pm3.94\%$ for the intermediate level bandwidths $\Delta\lambda = 20$, $30$, and $40$ nm, respectively. The decreasing value in the ANNs classification efficiency is a result of the fixed entanglement time of the photons. Note that by extending the band where intermediate states can be found, the probability of the photons (which possess a fixed spectral bandwidth) interacting with energy levels outside their spectra decreases. This leads to signals coming from samples with many levels to be wrongly classified as samples with one or two levels; thus affecting the identification efficiency of the ANN. However, we can overcome this by increasing the bandwidth of the photons, or equivalently by reducing the entanglement time. In the context of continuous-wave pumped SPDC sources, it has been demonstrated that one can produce photons with extremely short entanglement times, namely $T_e = 7.16$ fs \cite{Nasr2008}. This entanglement time corresponds to a bandwidth of 136 nm, which can perfectly fit our hypothetical samples. For the sake of completeness, we tested our algorithms for this entanglement time. The results of the improved mean efficiencies, and their corresponding standard deviations, are shown in the last two columns of Table \ref{tab:table1}. Note that all the mean efficiencies are above 99$\%$, precisely because the photon pairs overlap perfectly with the considered intermediate-level bands.

\section{Conclusion}

In summary, we have demonstrated a \emph{smart} protocol that is capable of extracting information about the number of intermediate levels that participate in the two-photon excitation of an arbitrary sample. Our proposed method makes us of artificial neural networks trained with various eTPA signals that are monitored/recorded as a function of an external delay between the absorbed photons. We have found that by selecting a proper entanglement time between photons, one can obtain classification efficiencies above 99$\%$, for groups of samples that contain up to four intermediate levels spread out among spectral bands of up to 40 nm, with a level resolution as high as 0.1 nm. Because of its high efficiency and low complexity, we believe that our proposed method shows potential for helping current experimental efforts that aim to demonstrate true eTPA-based imaging and spectroscopy.

\begin{acknowledgments}
\noindent This work was supported by DGAPA-UNAM under the UNAM-PAPIIT IN101623 project. The authors thank Mario A. Quiroz-Juárez for useful discussions.  
\end{acknowledgments}

\section*{Authors Declarations}

\subsection*{Conflict of interest}

\noindent The authors have no conflict of interest to disclose.

\subsection*{Authors contributions}

\noindent
Conceptualization, A. M. T. and R. J. L.-M.; formal analysis , A. M. T. and R. J. L.-M.; methodology A. M. T. and R. J. L.-M.; investigation, R. J. L.-M.; writing the manuscript, A. M. T. and R. J. L.-M.; project administration, A. M. T. and R. J. L.-M.
All authors contributed equally in this work. All authors have read and agreed to the published version of the manuscript.

\section*{Data Availability Statement}

The data that support the findings of this study are available from the
corresponding author upon reasonable request.

\nocite{*}
\bibliography{aipsamp}

\end{document}